\documentclass[10pt,notitlepage,a4paper,aps,prd,onecolumn,superscriptaddress,nofootinbib,groupedaddress]{revtex4-1}
\usepackage{array}

\usepackage[dvipsnames]{xcolor}

\usepackage[pdfencoding=auto, psdextra]{hyperref}
\usepackage[T1]{fontenc}
\usepackage[utf8]{inputenc}
\usepackage[english]{babel}
\usepackage{hyperref}
\usepackage[mathscr]{euscript}
\usepackage{caption}
\usepackage{subcaption}
\hypersetup{
colorlinks = true,
linkcolor = blue,
filecolor = magenta,
urlcolor = blue,
pdftitle = {TLSs interaction},
pdfpagemode = FullScreen,
citecolor = blue}
\usepackage{physics}

\usepackage{caption}

\usepackage{indentfirst}

\usepackage[pdftex]{graphicx}

\usepackage{amsmath}
\allowdisplaybreaks  
\usepackage{amssymb} 
\usepackage{amsfonts} 
\usepackage{enumerate}

\usepackage{fancyhdr}                     

\usepackage{float}                             

\usepackage{amsmath}         
\usepackage{amsfonts}        
\usepackage{amssymb}         
\usepackage{amsthm}          
\usepackage{empheq}          

\usepackage{color}           
\usepackage{ragged2e}        
\usepackage{titlesec}        

\usepackage{tipa}            

\usepackage[utf8]{inputenc} 
\allowdisplaybreaks          

\usepackage{enumitem}        

\usepackage{tikz}            
\usetikzlibrary{shapes.geometric,arrows.meta,arrows,positioning}  
\usetikzlibrary{tikzmark}   

\titleformat{\paragraph}
  {\normalfont\normalsize\bfseries}{\theparagraph}{1em}{} 
\titlespacing*{\paragraph}
  {0pt}{3.25ex plus 1ex minus .2ex}{1.5ex plus .2ex} 

\begin{document}
\title{Chiral environment effects on the dynamics of a central chiral molecule}

\author{Daniel Martínez-Gil}
\email{daniel.martinez@ua.es}
\affiliation{Fundacion Humanismo y Ciencia, Guzmán el Bueno, 66, 28015 Madrid, Spain.}
\affiliation{Departamento de F\'{\i}sica Aplicada, Universidad de Alicante, Campus de San Vicente del Raspeig, E-03690 Alicante, Spain.}

\author{Pedro Bargueño}
\email{pedro.bargueno@ua.es}
\affiliation{Departamento de F\'{\i}sica Aplicada, Universidad de Alicante, Campus de San Vicente del Raspeig, E-03690 Alicante, Spain.}

\author{Salvador Miret-Artés}
\email{s.miret@iff.csic.es}
\affiliation{Instituto de Física Fundamental, Consejo Superior de Investigaciones Científicas, Serrano 123, 28006, Madrid, Spain}

\begin{abstract}
In this work, we develop a system {\it plus} environment approach for interacting chiral molecules under a quantum-classical description of the spin-spin model using a Caldeira-Leggett-like coupling. After presenting the interacting model, we show that a long-ranged parity-nonconserving interaction, encoded within a non-linear Schrödinger equation, produces an energy difference between the two enantiomers of the central chiral molecule when it interacts with a chiral environment.  Three examples of such interactions are considered, with particular focus on $Z^0$-photon vacuum polarization. Finally, we reveal a {\it chirality transmission effect} phenomenon, where the time-averaged population difference of the central molecule is amplified when the chiral asymmetry of the environment is considered.
\end{abstract}
\maketitle

\section{Introduction}
Chiral molecules play a fundamental role in chemistry, physics, and biology, particularly due to their optical activity and asymmetric interactions in biological systems. Within a simple
approach, these molecules are commonly described using two-level systems or double-well potentials, where the two states correspond to the left- and right-handed enantiomers \cite{Harrisystodolsky, Quack1986}. In quantum mechanics, tunneling effect allows the interconversion between these enantiomers, introducing an intriguing challenge in understanding the stability of certain chiral molecules.

A well-known paradox in this context was introduced by Hund, who questioned why certain chiral molecules, such as amino acids and sugars, are mainly observed in only one enantiomeric form despite the possibility of quantum tunneling between the two states \cite{hund2} (this problem being usually known as molecular homochirality). This apparent discrepancy, often referred to as Hund's paradox, suggests the need for an additional mechanism that suppresses or prevents the tunneling process, leading to the stability of a single enantiomer.

One possible solution to Hund’s paradox was proposed by Harris and Stodolsky in 1978 \cite{Harrisystodolsky}. They introduced the concept of the parity-violating energy difference (PVED), which arises from weak interactions and leads to a small energy difference between the two enantiomers. This concept was previously suggested in earlier works \cite{lethokov, KOMPANETS1976414, ARIMONDO1977} and provides a physical basis for the observed chiral stability in nature.
By considering a regime where the PVED is much higher than the tunneling effect, the PVED ensures that one enantiomer becomes energetically favorable over the other, suppressing tunneling and stabilizing a single enantiomeric form \cite{Harrisystodolsky}. 

Despite extensive experimental efforts, direct detection of PVED remains challenging due to the extreme sensitivity required. Two of the most long-standing experimental efforts are under development in Zürich \cite{Quack1986}  and Paris \cite{Letokhov1975, Letokhov1976}.
Referring to the Zürich one, recent improvements have yielded a sensitivity of $\epsilon_{PV} = 100 \; aeV$ \cite{Dietiker}, which aligns with theoretical predictions for molecules with lighter nuclei \cite{zurich1, zurich2, zurich3}. In
 the Paris experiment, the PVED manifests as a frequency shift $\Delta \nu$ between enantiomers, given by:
\begin{equation}
    \epsilon_{PV}^*-  \epsilon_{PV}= h (\nu_R-\nu_L).
\end{equation}
The Paris group achieved measurements with a sensitivity of $\frac{\Delta \nu^{PV}}{\nu} \approx 10^{-13}$ \cite{Paris1999}, expecting that molecules containing Ruthenium and Osmium could present $\frac{\Delta \nu^{PV}}{\nu} \approx 10^{-14}$ \cite{experimentos}, bringing us closer to PVED observation.

Several recent experimental approaches are also under development. These include studies using trapped chiral molecular ions \cite{Landauetal}, rotational spectroscopy techniques predicting PVED-induced shifts on the order of $10^{-14}$ \cite{sahu2023detection}, and proposals for parity-violating precision measurements in racemic samples \cite{PhysRevX.13.041025}. Additionally, tailored microwave fields have been suggested to control quantum states of chiral molecules, providing a novel route for PVED detection \cite{lee2024quantum}. 

Although PVED has not been directly detected, theoretical calculations and experimental constraints indicate that its value is extremely small. Therefore, how such a small contribution can produce the molecular homochirality? 
To achieve a homochiral state, it is first necessary to generate an enantiomeric excess (attributed to PVED in this case) and subsequently amplify it. The intriguing hypothesis that PVED could be the initial source of enantiomeric excess, later amplified into the asymmetry observed in life, was first proposed by Ulbricht in 1959 \cite{Ulbricht1, Ulbricht2} and later revisited by Yamagata in 1966 \cite{Yamagata}.
During the 1980s, Kondepudi and Nelson \cite{Kondepudi} developed stochastic models based on a Frank-type autocatalytic model \cite{FRANK1953459}, allowing them to research the sensitivity of the Spontaneous Mirror Symmetry Breaking (SMSB) process to extremely weak chiral influences. Their calculations indicated that the energy values required to produce the SMSB process were about the estimated PVED range for biomolecules (the interested reader can see \cite{RevewHomochirality} as a review of biological homochirality).
Therefore, one possible solution to Hund's paradox is the PVED, which can be amplified by autocatalytic models.

Another possible solution to this suppression of the tunneling effect is the decoherence process, which is one of the most accepted solutions to Hund's paradox.
The decoherence process is a direct consequence of considering open quantum systems. In particular, it consists of blocking the tunneling effect due to the interaction between a system (a chiral molecule in our case) and an environment, which has been widely studied with different formalisms throughout history. 
In a pioneering work of Harris and Stodolsky \cite{Harris1981}, they conclude that ``collisions or interactions with the medium tend to stabilize rather than destroy the asymmetry of the system''. They use the density matrix formalism to introduce collisions between chiral molecules, which tends to stabilize them in the L or R state (see another related work of Simonius \cite{Simonius1978}). This idea has been reinforced for example in a more recent work of Trost and Hornberger \cite{Trost}, in which the authors used a collisional decoherence mechanism to stabilize states of chiral molecules. 

In this work, we want to combine the two previously mentioned effects which were crucial to explain Hund's paradox, namely the PVED and open quantum systems. In particular, 
we will use a system {\it plus} environment formalism considering a central chiral molecule as the system, and the environment as a collection of chiral molecules. We will use canonically conjugate variables to model a spin-spin problem \cite{weiss}, working within a Caldeira-Legget-like coupling \cite{caldeiraleggett1, Caldeiralegget2, CaldeiraLegget3, dani2}. 
 
This work is organized as follows: in Section \ref{section2}, we present examples of fundamental interactions capable of producing PVED in chiral molecules. In Section \ref{section3}, we introduce and motivate 
the system {\it plus} environment formalism we will use. In order to understand the interacting model we propose, Section \ref{section5} deals with possible long-range parity-nonconserving interactions which could be encoded within our model. In Section \ref{section6} we expose the main results of our interacting model.
Finally, Section \ref{section7} provides the conclusions of the present work.

\section{Fundamental interactions behind PVED}\label{section2}

To determine which interactions can induce PVED in chiral molecules, Barron introduced the fundamental concepts of true and false chirality \cite{barronfund}. True chirality characterizes systems that exist in two distinct enantiomeric states, which are interconverted by spatial inversion but not by time reversal combined with any proper spatial rotation \cite{barronpv}. Alternatively, false chirality describes systems where enantiomeric states can be interconverted by both spatial inversion and time reversal followed by spatial rotation. These distinctions are essential for understanding the origins of PVED, as only a truly chiral influence can generate this PVED in chiral molecules \cite{barronpv}. Barron's conclusions were based on the symmetry properties of quantum states describing chiral molecules and the interaction Hamiltonians under parity inversion ($\hat{P}$), time reversal ($\hat{T}$), and spatial rotations. 

\begin{figure}[h]
\centering
\includegraphics[width= 10cm]{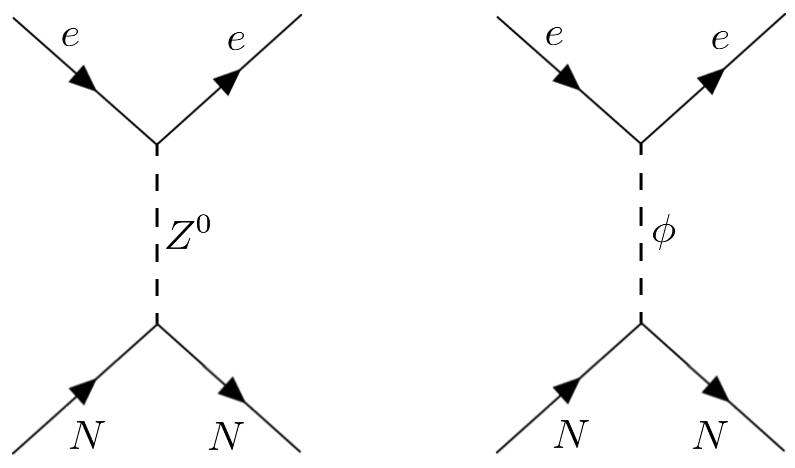}
\caption{\justifying Feynman diagrams of an electron-nucleon interaction. In the left panel, the interaction is mediated by the $Z^0$ while in the right panel by a scalar field (in this case an axion-like particle).}
\label{diagramaEWaxion}
\end{figure}

 In terms of fundamental symmetries, a Hamiltonian is truly chiral if it preserves both parity and time inversion symmetries, whereas it is falsely chiral if it preserves parity but violates time inversion. 
This distinction is crucial in the context of PVED, as only a truly chiral Hamiltonian can generate a PVED
in chiral molecules~\cite{barronpv}, a result that has been recently demonstrated by the authors within a quantum field theory framework \cite{Dani}. As key examples, weak neutral currents are truly chiral, while axions were proposed as representative cases of false chirality (see examples of these interactions in terms of Feynmann diagrams in Fig. \eqref{diagramaEWaxion}).

In Fig. \eqref{diagramaEWaxion} we depict the Feynman diagram showing the interaction between electrons and nucleons mediated by the $Z^0$ boson, which gives place to a 
truly chiral influence (for more details see, for instance, \cite{BERGER2004188} and references therein). The corresponding Hamiltonian is given by  

\begin{equation}\label{Hamiltonianoew}
H_{NC} = \frac{G_F}{4\sqrt{2}}\sum_i\sum_j Q_{W_i} \{ \Vec{p}_j \cdot \Vec{\sigma}_j,\rho(\Vec{r_i}-\Vec{r}_j)\},
\end{equation}  

where $G_F$ is the Fermi constant, $\Vec{p}$ represents the electron momentum, $\Vec{\sigma}$ is the electron spin, and $\rho$ is the nucleon density (with electrons and nucleons indexed by $j$ and $i$, respectively). The weak charge $ Q_W $ is given by  

\begin{equation}
Q_W = (1 - 4\sin^2\theta_W) Z - N,
\end{equation}  

where $\theta_W$ is the Weinberg angle, while $ Z $ and $ N $ correspond to the number of protons and neutrons, respectively. As a remark, the Hamiltonian \eqref{Hamiltonianoew} arises as a non-relativistic approximation within quantum field theory \cite{BERGER2004188}, and it can also be expressed in terms of the $\gamma^5$ Dirac matrix \cite{reviewBerger, axionesberger}  as 
\begin{equation}\label{Hamiltonianoewgamma5}
H_{NC} = \frac{G_F}{2\sqrt{2}}\sum_i\sum_j Q_{W_i} \gamma^5_j\rho(\Vec{r_i}-\Vec{r}_j),
\end{equation}

At this point, it is easy to confirm that this Hamiltonian is truly chiral. This property arises from the presence of the scalar product $ \Vec{\sigma}_e \cdot \Vec{p} $, where $ \Vec{\sigma}_e $ is a time-odd axial vector, and $ \Vec{p} $ is a time-odd polar vector. Therefore, the Hamiltonian transforms as a time-even pseudoscalar, violating $\hat{P}$ while preserving $\hat{T}$. As a result, it can be shown that it can generate PVED in chiral molecules (see \cite{Dani} for details within both quantum field theory and a non-relativistic framework).  

Alternatively, the same electron-nucleon interaction mediated by an axion, whose Feynman diagram is shown in the right panel of Fig. (\ref{diagramaEWaxion}), is falsely chiral. In natural units ($ c = \hbar = 1 $), the corresponding interaction Hamiltonian \cite{moody} takes the form  

\begin{equation}\label{axion}
H_{ax} = (g_s^N g_p^e) \frac{\Vec{\sigma}_e \cdot \Vec{r}}{8\pi m_e} \left(\frac{m_{\phi}}{r}+\frac{1}{r^2}\right) e^{-m_\phi r},
\end{equation}  

where $ g_s^N $ and $ g_p^e $ are the scalar and pseudoscalar coupling constants of the axion to the nucleon and electron, respectively, while $ \Vec{r} $ represents the separation vector between the electron and nucleon, and $ m_\phi $ is the axion mass.  

In the Hamiltonian \eqref{axion}, all terms are scalars except for $ \Vec{\sigma}_e $ (a time-odd axial vector) and $ \Vec{r} $ (a time-even polar vector). This structure causes the Hamiltonian to transform as a time-odd pseudoscalar, therefore violating both $\hat{P}$ and $\hat{T}$. Thus, it can not induce PVED in chiral molecules \cite{Dani}.  

\section{Simple model for chiral molecules within a quantum-classical formalism}\label{section3}

\subsection{Isolated model}

Although both electroweak and axion-like effects can be incorporated in quantum chemical calculations (citas de algo), here we employ a minimalistic two-level model to describe chiral molecules.
Within this approximation, the Hamiltonian of a chiral molecule considering a parity violation contribution is usually represented by \cite{Harrisystodolsky}
\begin{equation}\label{hamilsigmas}
    \hat{H} = \delta \hat{\sigma_x}+\epsilon\hat{\sigma_z},
\end{equation}
where $2\epsilon$ is the PVED, representing the energy difference between the left and the right enantiomers, and 2$\delta$ is the energy difference between the eigenstates of the Hamiltonian $\ket{+}, \ket{-}$, being inversely proportional to the tunneling time between enantiomers. In the presence of PVED, the eigenstates of the Hamiltonian can be expressed in terms of the localized states $\ket{L}$ and $\ket{R}$ as
\begin{equation}\label{AB}
    \begin{pmatrix}
\ket{+} \\
\ket{-}
\end{pmatrix}
=
\begin{pmatrix}
\sin\theta & \cos\theta\\
\cos\theta & -\sin\theta
\end{pmatrix}
    \begin{pmatrix}
\ket{L} \\
\ket{R}
\end{pmatrix},
\end{equation}
where $\theta$ is the mixing angle and can be obtained as
\begin{equation}
    \tan2\theta = \frac{2 H_{RL}}{H_{LL}-H_{RR}} = \frac{\delta}{\epsilon_{PV}}.
\end{equation}

With this in mind, the molecular wave function can be expressed as
\begin{equation}
    \ket{\psi(t)} = a_L(t)\ket{L}+a_R(t)\ket{R},
\end{equation}
the time-dependent Schrödinger equation being
\begin{align}
    i\hbar \dot{a}_L & = \epsilon a_L +\delta a_R, \\
    i\hbar \dot{a}_R & = \delta a_L -\epsilon a_R.
\end{align}

At this point, after performing a Madelung transformation, $a_{L, R}(t) = \abs{a_{L,R}(t)} e^{i\phi_{L,R}(t)}$,
to the Schrödinger equation, we obtain
\begin{align}
    \dot{z} &= -2\delta\sqrt{1-z^2} \sin \phi \label{z}\\
    \dot{\phi} &= 2\delta \frac{z}{\sqrt{1-z^2}}\cos \phi + 2\epsilon\label{phi},
\end{align}
where we have defined the population and phase differences as $z \equiv \abs{a_R(t)}^2-\abs{a_L(t)}^2$ and $\phi(t) \equiv \phi_L(t) - \phi_R(t)$, respectively \cite{PedroCPL2011}. The introduction of these variables, which act as  generalized momentum ($z$) and generalized position ($\phi$), makes possible to apply Hamilton equations to obtain the following Hamiltonian function 
\begin{equation}\label{hamil0}
    H_0 = -2\delta \sqrt{1-z^2} \cos \Phi + 2\epsilon z.
\end{equation}

We would like to remark that this quantum-classical description, which is motivated by the first Hopf fibration (the interested reader can see \cite{dani2} for more details on it), is similar to the 
Meyer-Miller-Stock-Thoss mapping \cite{Meyermiller, StockThoss}.

As our purpose in the present manuscript is to study the effects of an environment of chiral molecules, we need to go beyond the isolated model. 

\subsection{Environment as a collection of chiral molecules}

As in an usual system {\it plus} environment model, we will consider the following total Hamiltonian
\begin{equation}
    H_T = H_S+H_{E}+H_{I},
\end{equation}
where $H_S, H_E, H_I$ refer to the system, environment, and interaction respectively.

By employing a Caldeira-Legget-like formalism \cite{caldeiraleggett1, Caldeiralegget2, CaldeiraLegget3}, and neglecting the interaction between the chiral molecules of the environment, the total Hamiltonian 
can be written as
\begin{equation}
    H = \underbrace{-2\delta \sqrt{1-Z^2} \cos \Phi + 2\epsilon Z}_{H_S} + \underbrace{\sum_i \left[-2\delta_i\sqrt{1-z_i^2} \cos \phi_i + 2\epsilon_i z_i\right]}_{H_E}+\underbrace{Z\sum_i \Lambda_i z_i}_{H_I},
\end{equation}
where the index $i$ goes from 1 to $N$, being $N$ the total number of chiral molecules in the environment, and $\Lambda_i$ represents the coupling strength of the central chiral molecule with the $i$th chiral molecule of the environment. The corresponding Hamilton equations (2 for the system and $2N$ for the environment) read

\begin{eqnarray}
    \dot{Z} =& -2\delta\sqrt{1-Z^2}\sin \Phi, \label{Zsystem}\\
    \dot{\Phi} =& 2\epsilon+2\delta \frac{Z}{\sqrt{1-Z^2}}\cos \Phi + \sum_i \Lambda_i z_i,\\
    \dot{z}_i = & -2\delta_i \sqrt{1-z_i^2}\sin \phi_i,\\
    \dot{\phi}_i =& 2\epsilon_i+ 2\delta_i \frac{z_i}{\sqrt{1-z_i^2}} \cos \phi_i + Z\sum_i \Lambda_i.\label{phienvironment}
\end{eqnarray}

Due to the chaotic behaviour of the system for $N>1$ (please see \cite{dani2} for more details on it), an averaging of the population differences in the number of realizations, $n$, $\langle Z(t)\rangle_n$, is mandatory in order to obtain conclusions. 
\\
\\
Let us remark that the crucial feature we have introduced in our model is the $\Lambda$ parameter, which couples the environment to the central molecule. In the following section we will 
propose a physical realization of this coupling in terms of a long-range parity violating interaction.

\section{Nature of the interaction}\label{section5}
In order to interpret the interaction we have included in our model, we will recover the previously used variables $a_L(t), a_R(t)$ to obtain a Schrödinger-like equation, where the effects of the coupling become manifest and easily interpretable. We will consider that the wave function describing a single chiral molecule of the environment can be expressed as $\ket{\varphi} = b_{L_i}(t) \ket{L_i} + b_{R_i}(t)\ket{R_i}$. Therefore, the following change of variables

\begin{align*}
    Z &= \abs{a_R(t)}^2-\abs{a_L(t)}^2, \Phi(t) \equiv \Phi_L(t) - \Phi_R(t) \rightarrow & a_{L, R}(t) = \abs{a_{L,R}(t)} e^{i\Phi_{L,R}(t)}, \\
    z_i &= \abs{b_{R_i}(t)}^2-\abs{b_{L_i}(t)}^2, \phi_i(t) \equiv \phi_{L_i}(t) - \phi_{R_i}(t) \rightarrow & b_{L, R_i}(t) = \abs{b_{L,R_i}(t)} e^{i\phi_{L,R_i}(t)},
\end{align*}
lead to  
\begin{eqnarray}
i\hbar \dot{a}_L =& \epsilon a_L + \delta a_R + \frac{1}{2} a_L   \Lambda \sum_i (\abs{b_{R_i}}^2-\abs{b_{L_i}}^2), \\
i\hbar \dot{a}_R =& \delta a_L - \epsilon a_R - \frac{1}{2} a_R  \Lambda \sum_i (\abs{b_{R_i}}^2-\abs{b_{L_i}}^2), \\
i\hbar \dot{b}_{L_i} =& \epsilon_i b_{L_i} + \delta_i b_{R_i} + \frac{1}{2} b_{L_i}  N \Lambda (\abs{a_R}^2-\abs{a_L}^2), \\
i\hbar \dot{b}_{R_i} =& \delta_i b_{L_i} - \epsilon_1 b_{R_i} - \frac{1}{2} b_{R_i}  N \Lambda (\abs{a_R}^2-\abs{a_L}^2).
\end{eqnarray}

This is a non-linear Schrödinger equation (Gross-Pitaevskii-like), whose Hamiltonian is given by

{\small
\begin{equation}\label{Hamgrande}
\hat{H} =
\begin{array}{c c}
\hspace{-7.1cm}
\overbrace{\phantom{xxxxxxxxxxxx
xxxxxxxxxxxxxxxxxxxxxxxxxxxxxx}}^{\hat{H}_1} \\
\begin{pmatrix}
\epsilon  + \frac{1}{2} \Lambda \sum_i (\abs{b_{R_i}}^2-\abs{b_{L_i}}^2) & \delta & 0 & 0\\
\delta & - \epsilon - \frac{1}{2} \Lambda \sum_i(\abs{b_{R_i}}^2-\abs{b_{L_i}}^2) & 0 & 0\\
0 & 0 & \epsilon_i +  \frac{1}{2} \Lambda (\abs{a_R}^2-\abs{a_L}^2) & \delta_i\\
0 & 0 & \delta_i & -\epsilon_i -  \frac{1}{2} \Lambda (\abs{a_R}^2-\abs{a_L}^2)\\
\end{pmatrix}
\\
\hspace{+8.1cm}\underbrace{\phantom{xxxxxxxxxxxxxxxxxxxxxxxxxxxxxxxxxxxx}}_{\hat{H}_i}
\end{array}
\end{equation}}

 In order to obtain the energy of each state of the central chiral molecule, we have to diagonalize $\hat{H}$, which can be built by blocks in terms of $\hat{H}_1, \hat{H}_i$. Diagonalizing $\hat{H}_1$, we obtain the following eigenvalues:
\begin{equation}
    \lambda_{1,2} = \pm \sqrt{(\epsilon_1^2+\delta_1^2 + \epsilon_1 \Lambda \sum_i(\abs{b_{R_i}}^2-\abs{b_{L_i}}^2)+\frac{1}{4}\Lambda^2(\abs{b_{R_i}}^2-\abs{b_{L_i}}^2))}.
\end{equation}

We have to remind that the Hamiltonian $\hat{H}_1$ is written in the $\ket{+}, \ket{-}$ base ($E_{\pm} = \lambda_{1,2}$). Therefore, the energies of the $\ket{L_1}$ and $\ket{R_1}$ states are given by
\begin{eqnarray*}
    E_{L_1} = \bra{L_1}\hat{H_1}\ket{L_1} = & (\cos \theta \bra{+}+ \sin \theta \bra{-}) \hat{H}(\cos \theta\ket{+}+ \sin \theta \ket{-}) =E_+  \cos^2 \theta +  E_-\sin^2 \theta\\ 
    E_{R_1} = \bra{R_1}\hat{H_1}\ket{R_1} = & (\sin \theta \bra{+}- \cos \theta \bra{-}) \hat{H}(\sin \theta\ket{+}- \cos \theta \ket{-}) =E_+ \sin^2 \theta  + E_-\cos^2 \theta . 
\end{eqnarray*}

If we define the energy difference as $\Delta E_{L_1, R_1} = E_{L_1}-E_{R_1}$, we obtain
\begin{equation} \label{DifEn}
    \Delta E_{L_1, R_1} = 2 (\sin^2\theta - \cos^2\theta)\sqrt{\epsilon_1^2+\delta_1^2 + \epsilon_1 \Lambda \sum_i (\abs{b_{R_i}}^2-\abs{b_{L_i}}^2)+\frac{1}{4}\Lambda^2\sum_i(\abs{b_{R_i}}^2-\abs{b_{L_i}}^2)^2},
\end{equation}
which is one of the main results of this work.

From this expression it can be seen that the energy difference between the left and right states of the central molecule  now incorporates two extra terms: one coming from the $\epsilon_1$-$\Lambda$ coupling and other which goes with $\Lambda^2$. It is remarkable that these new two terms are non-zero when the population difference of the molecules of the environment is different from zero, which can be achieved by increasing the value of $\Lambda$ (see the next section for numerical results).

Following a similar procedure, we can obtain the energy difference between the left and right states of the molecules of the environment, which is given by
\begin{equation}
     \Delta E_{L_i, R_i} = 2 (\sin^2\theta - \cos^2\theta)\sqrt{\epsilon_i^2+\delta_i^2 + \epsilon_i \Lambda (\abs{a_R}^2-\abs{a_L}^2)+\frac{1}{4}\Lambda^2(\abs{a_R}^2-\abs{a_L}^2)^2},
\end{equation}
where the $\Lambda$ terms depend now on the population difference of the central chiral molecule.
\\
\\
Summarizing, we have shown that the $\Lambda$-term gives place to an energy difference between the left and right states of each chiral molecule. But, as we discuss in Section \ref{section2}, only a truly chiral interaction, such as electroweak interactions, can produce this energy difference between enantiomers \cite{barronpv,Dani}. However, electroweak interactions are of very short range, which precludes them to be the main P-odd ingredient when considering  intermolecular interactions between our central molecule and those of the environment. But, what about long-range P-odd interactions?
The necessity for a long-range P-odd interaction arises from the disparity between the typical molecular scales (approximately $10^{-10}$ m) and the range of the electroweak interactions 
(around $10^{-18}$ m).  Therefore, given that our model considers a central chiral molecule interacting with an environment of other chiral molecules, a contact electroweak interaction could hardly produce observable intermolecular effects. Consequently, it seems plausible to propose that the interaction responsible for the parity-odd behaviour must be long-ranged.

\subsubsection{Long-range parity-nonconserving interactions through the $\Lambda$-term}

\begin{figure}[h]
\centering
\includegraphics[width= 5cm]{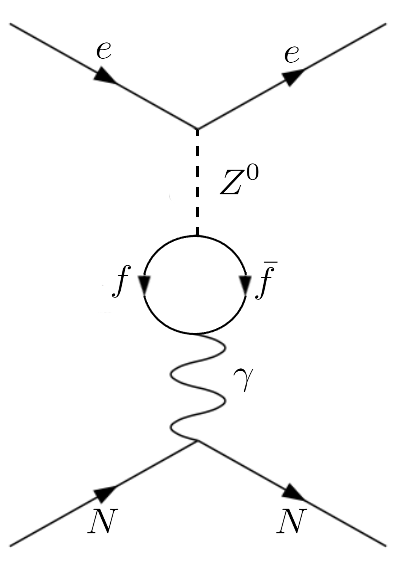}
\caption{\justifying Feynman diagram of an electron-nucleon interaction mediated by a mixed $Z^0-\gamma$ fermion vacuum polarization.}
\label{vacuum_polarization}
\end{figure}

 Up to our knowledge, there are three possible ways to make chiral molecules interact through a long-range parity-nonconserving interaction.
\begin{itemize}
\item i) P-odd Van der Waals forces. This idea was pioneered by Zhizhimov and Khriplovich \cite{PoddKhriplovich}, who discovered that the two-photon exchange between two atoms considering  parity nonconservation leads to a P-odd potential, which at long range decreases with distance as $r^{-8}$. Apart from the usual polarizability ($\alpha_{ik}$) used in the Van der Waals forces
\begin{equation}
    \alpha_{ik} = \sum_n \frac{\bra{0}d_i\ket{n}\bra{n}d_k\ket{0}+\bra{0}d_k\ket{n}\bra{n}d_i\ket{0}}{E_n-E_0},
\end{equation}
their potential depends  on the P-odd polarizability $\beta_{ik}$ (first introduced in \cite{Feinberg})
\begin{equation}
    \beta_{ik} = \sum_n \frac{\bra{0}d_i\ket{n}\bra{n}\mu_k\ket{0}+\bra{0}\mu_k\ket{n}\bra{n}d_i\ket{0}}{E_n-E_0}.
\end{equation}

As an aside note, we note that the authors estimated, using this mechanism, an energy difference of 100 Hz between the right- and left-handed helicoidal structure in rare-earth crystals \cite{CristalsKhriplovich, bookKhriplovich}. 

\item ii) Long-distance parity-nonconserving interaction between
charged particle and system. This idea, firstly introduced by Flambaum in \cite{Flambaum1}, considers that  
a charged particle creates electric ($\vec{E}$) and magnetic ($\vec{H}$) fields polarizing the system (atom, molecule, nucleus).
Using second-order perturbation theory, and imposing time invariance, the contribution of parity-nonconserving interaction to the energy can be written as
\begin{equation}\label{U}
    U = \beta_v (\vec{H} \times \vec{E}) \vec{J},
\end{equation}
where $\vec{J}$ and $\beta_v$ are the angular momentum and parity-nonconserving vector polarizability of the system. As the author stated, the leading long-range term in Eq. \eqref{U} goes as $r^{-4}$ if the particle has velocity or as $r^{-5}$ for small or zero velocity (e.g., for the
interaction of atomic ions in solids).

\item iii) Mixed $Z^0-\gamma$ fermion vacuum polarization. This idea was introduced by Flambaum and Shuryak in \cite{Flambaum3}. By studying the joint parity-violating characteristics of the $Z^0$ boson with the long-range of the photon  (see Fig. \eqref{vacuum_polarization}), they obtained the following parity-violating electron-nucleus interaction (considering the electron loop):
\begin{equation}\label{vacuum_potential}
    W = \frac{G}{2\sqrt{2}} \gamma_5\left[-Q_W\rho(r) + Z\frac{2\alpha (1-4\sin^2\theta_W)m_e^2}{3\pi^2r}I(r)\right],
\end{equation}
where $I(r)$ can be expressed as
\begin{equation}
    I(r) = \int_1^\infty \exp(-2xm_er)\sqrt{x^2 -1} \left(1+\frac{1}{2x^2}\right)dx.
\end{equation}

In Eq. \eqref{vacuum_potential}, the first term refers to the usual parity-violating electron nucleus interaction, and the second term is the long-range contribution. $G$ is the Fermi constant, $Q_W \approx -N + Z (1-4\sin^2\theta_\omega)$ is the weak charge , where $Z$ and $N$ are the numbers of protons and neutrons and $\theta_{W}$ is the Weinberg angle. $\gamma_5$ is the Dirac matrix, $\rho(r)$ is the nuclear density and $m_e$ the electron mass. 

\end{itemize}

At this point, we would like to remark that mixed $Z^0-\gamma$ vacuum polarization is the only truly chiral interaction among all the long-range P-odd contributions previously considered. Therefore, it would
constitute the only possibility to create PVED between the two enantiomeric states \cite{barronfund, Dani}. Interestingly, in Ref. \cite{Flambaum2} , the authors pointed out that the aforementioned P-odd potential contributes to
parity-nonconserving effects in atoms and molecules through a radiative correction to the weak charge, which is the main source of the contact P-odd interaction usually considered within the molecular context.
In the next section, we will briefly explore the effects of this contribution within our simple model of a central (non-chiral) molecule immersed in a chiral environment.

\section{Results}\label{section6}

As was mentioned in \cite{Flambaum2}, the long-range term in Eq. \eqref{vacuum_potential} depends on the inverse of the fermion mass of the loop. For example, if the fermion is an electron, it exceeds by 5 orders of magnitude the typical electroweak range. Although the electron loop only contributes a 0.1$\%$ to the weak charge $Q_W$ (reaching 2\% when considering sum over all kind of fermions), the long-range nature of the interaction
could give place to very interesting results within chiral molecules.

As was discussed in the previous section, the interaction between the central molecule and those of the environment can be captured by the $\Lambda$ parameter, which represents a P-odd intermolecular effect. In this section, we will explore the effects of a chiral environment on the dynamics of the central molecule.  In particular, among the different results we have obtained (see \cite{dani2} for details), in this manuscript we are interested in what we refer to {\it chirality transmission effect} from the environment to the central molecule. 

\begin{figure}[h]
\centering
\includegraphics[width= 12cm]{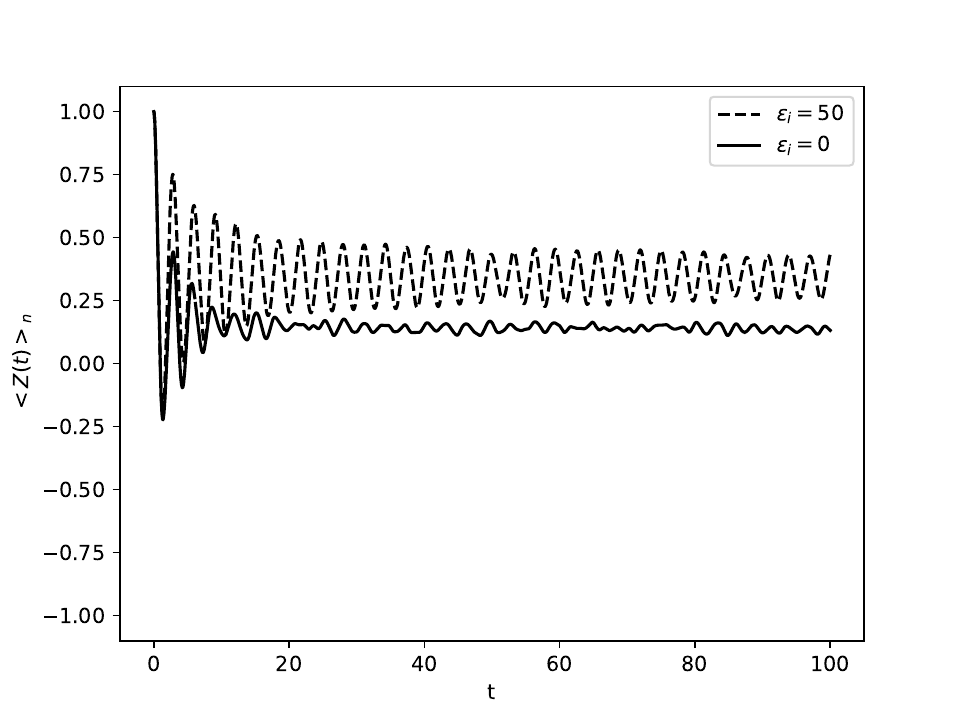}
\caption{ \justifying Averaged population difference of the system in the number of realizations $n$, considering $n = 2000$, $\Lambda = 1$, $\epsilon = 0$,  $\delta = \delta_i = 1$, and  $z_i(0)$ and $\phi_i(0)$ as random parameters between $(-1,1)$ and $\left[0, 2\pi\right)$ respectively. The dashed line represents $\epsilon_i = 50$, and the solid line $\epsilon_i = 0$.}
\label{figura11}
\end{figure}

We remind the reader that, due to the chaotic behaviour of the coupled equations, we must employ averages of the population difference in the number of realizations $n$ ($\langle Z(t)\rangle_n$).
Before discussing the results shown in Fig. \eqref{figura11}, it is important to note that if we set \(\epsilon = \epsilon_i = \Lambda = 0\), the system would exhibit symmetric oscillations between the L and R states, yielding a vanishing time-averaged population difference. As shown in Fig. \eqref{figura11}, we consider two cases, both with $\epsilon = 0$ and $\Lambda = 1$. In the case represented by the solid line,  chiral molecules in the environment are assumed to be symmetric ($\epsilon_i = 0$). Here, the time-averaged population difference is different from 0 ($\langle Z(t)\rangle_t \approx 0.12$), and a clear damping effect is observed, only because $\Lambda \neq 0$. In contrast, the dashed line corresponds to the case where a nonzero PVED is introduced in the environment ($\epsilon_i \neq 0$). In this situation, we obtain $\langle Z(t)\rangle_t \approx 0.30$. The increase from 0.12 to 0.30 is exclusively due to a nonzero PVED in the environment. This phenomenon, which we refer to as the {\it chirality transmission effect}, shows a clear enantioselection effect in the central chiral molecule due to the asymmetry of the environment.

Finally, we would like to note that the parameters employed in Fig. \eqref{figura11} are compatible with those predicted by the theory. On one hand, we have
chosen the short-ranged PVED, $\epsilon_{i}=50$, and the long-ranged one, $\Lambda=1$, in fully agreement with the predictions of \cite{Flambaum2} for the contribution of Fig. \ref{vacuum_polarization} regarding
the $Z_0$-photon vacuum polarization case. On the other hand, for the molecules of the environment we have chosen the relation $\delta_i = 0.02 \epsilon_i$, which is roughly the same obtained by employing high-level {\it ab-initio} calculations for certain chiral molecules (see the data table provided in \cite{quack2008}). We remind the reader that the central chiral molecule fulfills  $\delta \gg \epsilon$, which can be easily satisfied for several candidates \cite{quack2008}.

\section{Conclusions}\label{section7}

In this work we have developed a system {\it plus} environment model for interacting chiral molecules, where environmental long-ranged P-odd effects 
produce an energy difference between the two enantiomers of the central chiral molecule. In addition, we have shown a {\it chirality transmission effect} 
between the environment and a central molecule under certain circumstances.

After reviewing the definitions of true and false chirality we introduce some P-odd Hamiltonians, some of them producing parity-violating energy differences between chiral molecules, 
emphasizing electroweak effects as truly chiral and axion-mediated as falsely chiral interactions. 

Our model is built by treating chiral molecules as simple two-level systems which can be
described, after performing a Madelung transformation, by a classical-like Hamiltonian.  By considering this classical-like formalism, we developed a system {\it plus} environment model for chiral molecules by 
coupling the population differences of the central chiral molecule with those of the environment, which corresponds to a Gross-Pitaevskii-like equation. 

This interacting model gives place to interesting effects. First, note that the system-environment coupling constant produces both a damping effect in the population difference of the central chiral molecule (therefore suppressing the tunnelling) and, importantly, an energy difference between the two enantiomers of it. Second, the presence of a nonzero parity-violating energy difference in the environmental chiral molecules induces an enantioselective effect on the central molecule, which we referred to as {\it chirality transmission effect.}

Furthermore, we explored the interpretation of the introduced interaction in terms of long-range parity-nonconserving effects by showing some possibilities within the definitions of true and false chirality, 
concluding that our model can be thought to include $Z^0$-photon mixed vacuum polarization effects.

\section*{Acknowledgements}
D. M. -G. acknowledges Fundación Humanismo y Ciencia for financial support. D. M. -G. and P. B. acknowledge Generalitat Valenciana through PROMETEO PROJECT CIPROM/2022/13.

\bibliographystyle{unsrt}
\bibliography{referencias.bib}

\end{document}